\documentclass{ptapap}
\usepackage{relsize}
\usepackage{url}

\author{Gianfranco De Zotti}[OAPd]
\author{Matteo Bonato}[IRA,OAPd]
\author{Zhen-Yi Cai}[CAS,SASS]
\affil[OAPd]{INAF-Osservatorio Astronomico di Padova, \\ Vicolo dell'Osservatorio 5, I-35122 Padova, Italy
}
\affil[IRA]{INAF, Osservatorio di Radioastronomia, Via Gobetti 101, I-40129, Bologna}
\affil[CAS]{CAS Key Laboratory for Research in Galaxies and Cosmology, \\ Department of Astronomy,
University of Science and Technology of China, \\ Hefei, Anhui 230026, China}
\affil[SASS]{School of Astronomy and Space Science, University of Science and Technology of China, Hefei 230026, China}

\title{Star formation across cosmic time with radio surveys. The promise of the SKA}
\headtitle{Star formation across cosmic time}
\begin{document}

\maketitle

\def\simlt{\mathrel{\rlap{\lower 3pt\hbox{$\sim$}}\raise 2.0pt\hbox{$<$}}}
\def\simgt{\mathrel{\rlap{\lower 3pt\hbox{$\sim$}} \raise
2.0pt\hbox{$>$}}}
\def\lsim{\,\lower2truept\hbox{${<\atop\hbox{\raise4truept\hbox{$\sim$}}}$}\,}
\def\gsim{\,\lower2truept\hbox{${> \atop\hbox{\raise4truept\hbox{$\sim$}}}$}\,}

\begin{abstract}
This lecture briefly reviews the major recent advances in radio astronomy
made possible by ultra-deep surveys, reaching microJansky flux density
levels. A giant step forward in many fields, including the study of the
evolution of the cosmic star formation history is expected with the advent of
the Square Kilometer Array (SKA).

\end{abstract}

\section{Introduction}

Ever since radio surveys have reached sub-milliJansky\footnote{One Jansky (Jy)
corresponds to $10^{-23}\,\hbox{erg}\,\hbox{cm}^{-2}\,\hbox{s}^{-1}\,
\hbox{Hz}^{-1} = 10^{-26}\,\hbox{W}\,\hbox{m}^{-2}\,\hbox{Hz}^{-1}$. }  flux
density levels, they proved to be a primary means of identifying star-forming
galaxies at high redshift. Observations have solidly demonstrated a tight
correlation between the low-frequency radio continuum, dominated by synchrotron
emission of relativistic electrons mostly produced by supernovae, and
far-infrared (FIR) emission of galaxies.

Since the FIR is a well established measure of the star-formation rate (SFR) of
galaxies, this correlation has legitimated the radio continuum emission as a
SFR tracer. However so far  radio surveys have only played a marginal role in
the study the cosmic star-formation history, mostly because of their
sensitivity. But the radio astronomy is at the verge of a revolution, thanks to
the square kilometre array (SKA), that will offer an observing window between
50\,MHz and 24\,GHz, extending to flux density limits more than three orders of
magnitude deeper than it is currently possible, with unprecedented versatility.

This lecture briefly reviews the role of radio astronomical surveys towards our
understanding of extragalactic sources. It focusses in particular on
populations of faint ($\mu$Jy) radio sources and discusses the advances
expected from the future deeper surveys with the SKA and its precursors, and
their impact on our understanding of the cosmic star-formation history.

The layout of the lecture is the following. Section~\ref{sect:introduction}
contains a general introduction to the radio sky.
Section~\ref{sect:AGNgeneralities} presents general properties of classical
radio sources, i.e. radio Active Galactic Nuclei (AGNs).
Section~\ref{sect:SFG} discusses the radio emission from star-forming galaxies.
The identification of the faint (sub-mJy) population is dealt with in
Section~\ref{sect:identifications}. Section~\ref{sect:RQ} deals with radio
quiet (RQ) AGNs, that were proposed as a new faint radio source population.
Section~\ref{sect:SKA} provides an introduction to the SKA and to its role in
the study of the cosmic star-formation history. Finally,
Section~\ref{sect:conclusions} summarizes the main conclusions.

\section{Introduction to the radio sky}\label{sect:introduction}

The radio sky is very different from what we are used to see in the optical.
With few exceptions (the nearest galaxies like Andromeda or, in the southern
hemisphere, the Magellanic Clouds, and solar system objects) the brightest
optical sources, visible with the naked eye, are stars. These are approximate
blackbody radiators (thermal emitters). Their emission covers a relatively
narrow range of frequencies, ranging from the ultraviolet (UV) to the infrared
(IR), depending on the star temperature. As one goes fainter, at optical
magnitudes $m\ge 19$--20 (in the AB system), galaxies take over. Most of their
light is again of thermal origin, coming from stars, so is confined at UV to IR
wavelengths.

But few stars are visible at radio frequencies. Most of the point sources
visible on radio maps are distant \citep[median redshift $z \sim 0.8$ for a
broad range of detection limits;][]{Condon1989} luminous radio galaxies (RGs)
and quasars. The radio emission from RGs and quasars is non-thermal. It  is due
to ultra-relativistic ($E\gg m_e c^2$) electrons moving in a magnetic field and
thereby emitting synchrotron radiation, which, unlike blackbody emission, can
cover a very large range in frequency, reaching up to $\sim 10$ decades in some
sources.

The fact that bright radio sources are mostly at substantial redshifts has
implied that, for several decades, extragalactic radio surveys remained the
most powerful tool to probe the distant universe. Even "shallow" radio surveys,
those of limited radio sensitivity, reach sources with redshifts predominantly
above 0.5.

Since the 1960's, the most effective method for finding high-$z$ galaxies has
been the optical identification of radio sources, a situation persisting until
the mid-1990's, when the arrival of the new generation of 8--10\,m class
optical/infrared telescopes, the refurbishment of the Hubble Space Telescope,
the Lyman-break technique \citep{Steidel1996} and the Sloan Digital Sky Survey
\citep[SDSS;][]{York2000} produced an explosion of data on high-redshift
galaxies.

Indeed radio surveys produced real revolutions in astrophysics and cosmology
\citep{DeZotti2010} :
\begin{itemize}

\item \textbf{Active Galactic Nuclei.} Radio emission provided the first
    evidence of non-stellar activity. Identifications of some of the sources
    detected by the earliest radio astronomical observations
    \citep{Bolton1949, Ryle1950} showed that they are extragalactic
    \citep{Ryle1950, BaadeMinkowski1954}. The amount of energy in the
    radiating particles turned out to be huge, up to $10^{60}\,$erg, for
    double-lobe radio galaxies \citep{Burbidge1959}, and difficult to account
    for.

\item \textbf{Synchrotron emission.} Estimated brightness temperatures of
    radio sources could not be reconciled with thermal emission from
    interstellar gas. This led to the identification of synchrotron emission
    as the dominant continuum process producing the power-law spectra of
    radio sources \citep[][]{AlfvenHerlofson1950, Shklovskii1952}; see
    \citet{GinzburgSyrovatskii1965} for an early review.

\item Radio astronomy first \textbf{pushed the boundary of observable
    universe to cosmological distances}. \citet{RyleScheuer1955} argued that
    the isotropic distribution of ``radio stars'' placed the bulk of them
    beyond 50\,Mpc. When arcminute positional accuracy became available
    \citep{Smith1952} it was quickly realized that the majority of the host
    galaxies were beyond the reach of the optical telescopes of the epoch.
    \citet{Minkowski1960} measured a redshift of 0.46 for 3C295, the
    fifteenth brightest radio source at high Galactic latitude. This
    measurement remained the redshift record for a galaxy for 10 years.
    Efforts to identify and measure redshifts of bright radio sources led to
    the discovery that they are frequently located in rich clusters of
    galaxies. This was also the case for 3C295; its redshift measurement led
    to the discovery of the most distant cluster known at the time. In 1965,
    the redshift record was 2.012 for the radio quasar 3C9
    \citep{Schmidt1965}. Only after the turn of the century did the redshift
    record become routinely set by objects discovered in surveys other than
    at radio wavelengths \citep[e.g.,][]{Stern2000}.

\item \textbf{The discovery of quasars}, starting with 3C273
    \citep{Hazard1963, Schmidt1963}. Their extreme luminosities, 10 to 30
    times higher than those of the brightest giant ellipticals, coming from
    very compact regions, called for \textit{a new energy source}.
    \citet{HoyleFowler1963b} were the first to argue that ``only through the
    contraction of a mass of $10^7- 10^8\,M_\odot$ to the relativistic limit
    can the energies of the strongest sources be obtained.''  Soon after,
    \citet{Salpeter1964} showed that accretion into a ``Schwarzschild
    singularity'' can release an energy of $0.057\,c^2$ per unit mass, and
    thus can account for the quasar luminosity. A similar view was
    independently proposed by \citet{Zeldovich1964}. The idea that the quasar
    luminosity comes from accretion onto a super-massive black hole gained
    momentum when \citet{LyndenBell1969} argued that ``collapsed bodies''
    (i.e. black holes) should be common in galactic nuclei. However, the
    alternative view of non-cosmological redshifts, was also held by some
    groups. For example \citet{HoyleBurbidge1966} argued that quasars are
    coherent objects, ejected by galactic nuclei at relativistic speed. An
    argument in favour of a local origin of these objects was that
    relativistic electrons emitting optical and infrared synchrotron
    radiation would also boost up the energy of ambient photons by inverse
    Compton scattering. Repeated scatterings boost dramatically the inverse
    Compton luminosity, yielding a divergence called the ``inverse Compton
    catastrophe''. To avoid that Compton energy losses exceed the synchrotron
    losses, the magnetic field must be very strong, to the point that the
    lifetimes of electrons become extremely short, of order of seconds,
    making the cosmological interpretation of redshifts hardly tenable
    \citep{Hoyle1966}.

\item \textbf{Relativistic beaming.} \citet{Woltjer1966} pointed out that the
    \citet{Hoyle1966} argument depends strongly on the assumed isotropy of
    relativistic electrons. If, as is dynamically more likely, electrons flow
    in a narrow cone (jet) along the field lines of a radial field, the
    difficulty is eliminated. \textit{The discovery of super-luminal motions}
    of quasar radio components \citep{Cohen1971, Whitney1971} strongly
    supported the relativistic jet model. In fact \citet{Rees1966} had
    predicted, five years before the discovery, that relativistic bulk
    motions at small angles with respect to the line-of-sight can produce
    apparent super-luminal velocities. This led to the \textit{development of
    unified models} of radio sources: quasars and radio galaxies are one and
    the same, with orientation of the axis to the viewer’s line of sight
    determining classification via observational appearance
    \citep{AntonucciMiller1985, Barthel1989, UrryPadovani1995}.

\item \textbf{History of the Universe}. Radio surveys highlighted a new
    global property of the universe: \textit{cosmological evolution},
    disproving the then widely accepted Steady-State model. The debate
    generated by early radio surveys was passionate. The Steady-State versus
    Big Bang controversy was rooted in the simplest statistics to be derived
    from any survey: the integral source counts, the number of objects per
    unit sky area above given intensities or flux densities. As discussed by
    \citet{Ryle1955} and \citet{RyleScheuer1955}, the slope of such counts
    must be flatter than the Euclidean slope\footnote{In an Euclidean
    universe the number of sources $N$ is proportional to the volume, i.e. to
    $r^3$ for a sphere; the flux density, $S$, is $\propto r^{-2}$ so that
    $N\propto S^{-3/2}$.} of $-3/2$ in a Steady-State universe and in any
    reasonable Friedman model, if the source spatial distribution is
    independent of distance. But the source count from the 2C(ambridge) radio
    survey \citep{Shakeshaft1955} showed an integral slope of approximately
    $-3$. \citet{Ryle1955} and \citet{RyleScheuer1955} interpreted the 2C
    excess of faint sources in terms of the radio sources having far greater
    space density at earlier epochs and argued against the Steady-State
    theory. However, observations by \citet{MillsSlee1957} with the Sydney
    antenna did not confirm the steep 2C slope. These authors showed that
    confusion, i.e. the blending of weak, individually undetectable sources,
    to produce detectable signals, can have disastrous effects, and this
    affected the early Cambridge source counts. Still \citet{Mills1958},
    after correcting for source confusion and instrumental effects, found a
    slope of $-1.65$, steeper than the Euclidean slope although substantially
    flatter than the 2C slope. \citet{Scheuer1957} developed the $P(D)$
    technique, circumventing the confusion problem and showing that the
    interferometer results of 2C were consistent with the Sydney findings.
    \citet{RyleClarke1961} argued that observations of radio sources ``appear
    to provide conclusive evidence against the steady-state model''. But the
    damage had been done: some cosmologists, led by Hoyle, believed that
    radio astronomers did not know how to interpret their data.

\item The \textbf{discovery of the Cosmic Microwave Background} (CMB) in 1965
    constituted the definitive consecration of the Big Bang model
    \citep{PenziasWilson1965, Dicke1965}, thus making the `source-count
    controversy' irrelevant in one sense. The initial slope of the
    low-frequency integral radio source counts was finally assessed at $-1.8$
    \citep{Ryle1968}, confirming that Ryle's interpretation was right. While
    the discovery of CMB may indeed have shown that a Big Bang took place,
    the source counts demonstrated further that {\it objects in the Universe
    evolve either individually or as a population} -- a concept not fully
    accepted by the astronomy community until both galaxy sizes and
    star-formation rates were shown to change with epoch.

\item \textbf{Discovery of pulsars} \citep{Hewish1968}. In 1974 Hewish was
    awarded the Nobel prize in physics. Observations were done by his student
    Jocelyn Bell.

\item \textbf{Discovery of the first pulsar in a binary system
    \citep{HulseTaylor1975}, allowing tests of general relativity in a strong
    gravitational field}. Such a compact system with rapidly orbiting masses
    would radiate large quantities of gravitational radiation. Taylor
    monitored the pulsar system for decades, finding that the shift of the
    periastron time is in close agreement with predictions from general
    relativity. Russell Hulse and Joseph Taylor were awarded the Nobel Prize
    1993.

\item \textbf{Detection of primordial CMB anisotropies} \citep{Smoot1992},
    shedding light on the origin of structure in the universe and opening the
    way to ``precision cosmology''.

\end{itemize}
For a more comprehensive list of the key discoveries from radio astronomy in
the metre and centimetre wavebands see \citet{Wilkinson2004}. The latter
authors noted that ``radio telescopes have set a large part of the current
astronomical agenda'', that ``the largest radio telescopes of their day have
dominated the discoveries'' and that ``most scientific advances follow
technical innovation''. Hence they conclude that the SKA, with its performances
orders of magnitude better than those of present day radio instrumentation,
brims with promises of transforming our view of the universe.

\section{Generalities on radio AGNs}\label{sect:AGNgeneralities}

Source counts are currently recognized as essential data in delineating the
different radio-source populations and in defining the cosmology of AGNs. These
counts are dominated down to milli-Jansky (mJy) levels by the canonical radio
sources, believed to be powered by super-massive black-holes
\citep[e.g.,][]{Begelman1984} in galactic nuclei.

By today’s standards strong radio sources have flux densities $S\ge 1\,$Jy,
intermediate ones go down $\simeq 1\,\hbox{mJy}$, and the deepest surveys reach
$\mu$Jy levels \citep[for reviews, see][]{DeZotti2010, Padovani2016,
Padovani2017}.

Radio-source spectra are usually described as power laws ($S_\nu \propto
\nu^{\alpha}$)\footnote{We note that this convention is not universal; some
people still prefer the old `Cambridge' convention $S_\nu \propto
\nu^{-\alpha}$.}. AGN-powered radio sources are traditionally classified in two
main categories: steep- ($\alpha < -0.5$) and flat-spectrum ($\alpha >- 0.5$).
The early low-frequency (meter to decimeter wavelength) surveys, found radio
sources almost exclusively of steep power-law form. Later surveys at
cm-wavelengths found objects of diverse spectral types, and in general,
anything which was not `steep-spectrum' in form was called `flat-spectrum'.
However, truly flat-spectrum sources are rare.

From a physical point of view the flat-/steep-spectrum classification based on
the low-frequency (typically $\simeq 1$ to $\simeq 5\,$GHz) spectral index
corresponds to the source compactness. Steep-spectrum sources are generally
extended while the flat-spectrum ones are compact. In fact, in a compactness
versus low frequency spectral index diagram there is a remarkably clean
separation between steep-spectrum/extended sources and flat-spectrum/compact
sources \citep[cf. Fig.~4 of][]{Massardi2011}. There are however exceptions,
such as the Compact Steep Spectrum (CSS) sources \citep[e.g.,][]{Orienti2016}.

Steep-spectra are optically thin synchrotron, while the ``flat-spectra'' are
the superposition of multiple components, with different self-absorption
frequencies. In general, however, each source has both a compact, flat-spectrum
core and extended steep-spectrum lobes \citep[cf., e.g., Fig.~2
of][]{DeZotti2010}. This already implies that a simple power-law representation
can only apply to a limited frequency range. Self-absorption (synchrotron and
free-free) can produce spectra rising with frequency at the low-frequency
optically-thick regime, while at high frequencies the synchrotron emission
becomes optically thin, power law, and energy losses of relativistic electrons
(``electron ageing effect'') translate into a spectral steepening.

According to the widely accepted ``unification'' scheme
\citep{ScheuerReadhead1979, OrrBrowne1982, Scheuer1987, Barthel1989}, the
appearance of sources depends primarily on their axis orientation relative to
the observer. This paradigm stems from the discovery of relativistic jets
\citep{Cohen1971, Moffet1972} giving rise to strongly anisotropic emission. A
line of sight close to the source jet axis offers a view of the compact,
Doppler-boosted, flat-spectrum base of the approaching jet. Doppler-boosted
low-radio-power \citet{FanaroffRiley1974}, type I (FR\,I; edge-dimmed)] sources
are associated with BL Lac objects, characterized by optically featureless
continua, while the powerful type II (FR\,II; edge-brightened) sources are seen
as flat-spectrum radio quasars (FSRQs). A third class of, mostly
steep-spectrum, radio sources, FR\,0, has been proposed by \citet{Baldi2015}.
These objects share many characteristics typical of FR\,I radio galaxies but
have a more conspicuous radio core and extend a few kpc at most.

The view down the axis offers an unobstructed sight of the active nucleus which
may outshine the starlight of the galaxy by five magnitudes. The source appears
stellar, either as a FSRQ or as a BL Lac object. FSRQs and BL Lacs are
collectively referred to as blazars. In the case of a side-on view, the
observed low-frequency emission is dominated by the extended, optically thin,
steep-spectrum components, the radio lobes; and the optical counterpart
generally appears as an elliptical galaxy. A dusty torus
\citep{AntonucciMiller1985} hides the active nucleus emission from our sight.
At intermediate angles between the line of sight and the jet axis, angles at
which we can see into the torus, but the alignment is not good enough to see
the Doppler-boosted jet bases, the object appears as a ‘steep-spectrum quasar’.


\begin{figure}
\vskip-0.85cm
\begin{center}
\includegraphics[width=0.7\textwidth]{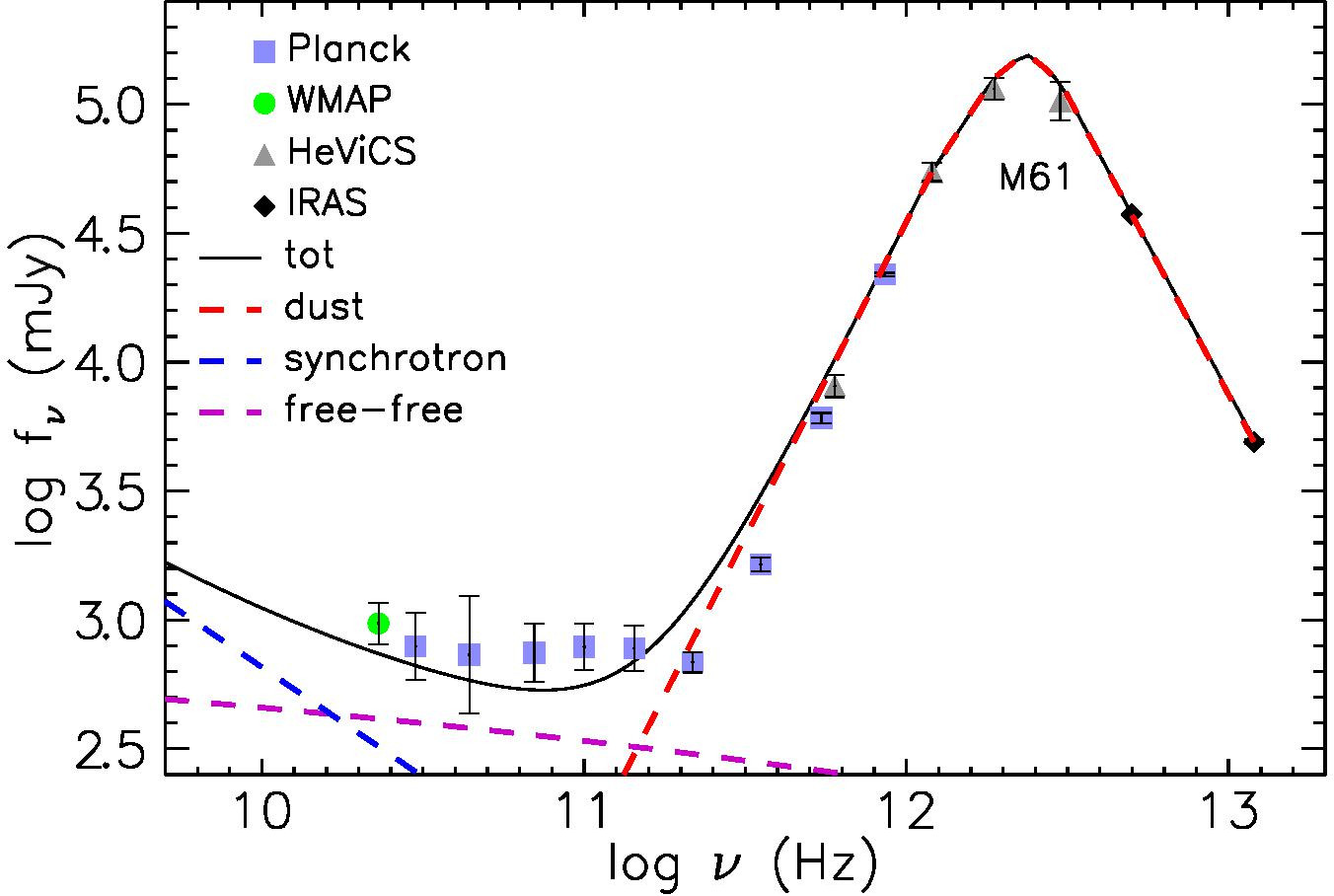}
\end{center}
\vskip-0.6cm
\caption{Spectral energy distribution (SED) of the nearby star-forming galaxy M\,61
showing the contributions of synchrotron (dashed blue line), free-free (dashed purple line)
and dust (dashed red line) emissions. The solid black line shows the total. Sources of data points:
\textit{Planck} \citep[open diamonds;][]{PCCS2}; WMAP \citep[filled circle][]{Gold2011};
the \textit{Herschel} Virgo Cluster Survey \citep[HeViCS, triangles;][]{Pappalardo2015} and IRAS (diamonds).
Note that photometric measurements of M\,61 are complicated because it is quite extended
(angular size of $6^\prime$, in the optical). On one side, the contributions of low surface density external regions
are easily missed; on the other side, the low resolution \textit{Planck} and \textit{WMAP} measurements may pick
up also contributions from our own Galaxy. Hence the true photometric errors may be substantially larger than the
nominal ones shown in the figure. The lines are not fits to the data but are simply meant to
illustrate the shape and the relative importance of the three contributions.
 } \label{fig:SED_SFG}
\end{figure}

\section{Radio emission from star-forming galaxies}\label{sect:SFG}

Counts down to mJy levels are accounted for by radio AGNs \citep[cf. Fig.~3
of][]{Massardi2010}. As mentioned above, the shape of the counts demonstrates
that these objects undergo a strong cosmological evolution. But when radio
surveys reached sub-mJy flux density levels \citep{Windhorst1984,
CondonMitchell1984}, the Euclidean normalized counts showed a flattening or an
upturn, indicating the emergence of a new population \citep[for updated counts
at 1.4\,GHz see][]{Padovani2016}.

\citet{Windhorst1985}, based on optical identifications available for less than
half of the sample, suggested that ``for $1 < S_{1.4} < 10\,$mJy a blue radio
galaxy population becomes increasingly important; these often have peculiar
optical morphology indicative of interacting or merging galaxies''. The
identification by \citet{Kron1985} showed that ``For $18.5 \le V \le 21.5$,
about one-third  of the mJy radio galaxies are bluer than the giant
ellipticals. They are not morphologically like the brighter  spirals, but
rather are a different class of peculiar (interacting, merging, or compact)
galaxies''.

Only a few years earlier, in 1983, the InfraRed Astronomical Satellite (IRAS)
had carried out an all sky survey at far-IR wavelengths, providing evidence of
a galaxy population with star-formation rates much higher than normal  spiral
galaxies (starburst galaxies). \citet{Danese1987} showed that evolving
starburst/interacting galaxies could easily account for the excess counts, for
the observed colours and for the morphological properties. On the contrary, the
data could not be interpreted in terms of a new population of unevolving
low-luminosity radio sources or of normal spiral galaxies with evolution
consistent with the observational constraints.

Although to really understand which sources were responsible for the flattening
and sort out the source population of the sub-mJy radio sky took more than
thirty years, it was already clear that the deep surveys were beginning to
detect radio emission not of nuclear origin but associated to star-formation.

Determinations of the local radio luminosity function \citep{Sadler2002,
MauchSadler2007} have shown that it has contributions from two populations. At
1.4\,GHz the bright portion ($P_{1.4\rm GHz}\ge
10^{23}\,\hbox{W}\,\hbox{Hz}^{-1}$) is dominated by canonical, steep-spectrum
radio galaxies, whose luminosities reach $P_{1.4\rm GHz}\ge
10^{26}\,\hbox{W}\,\hbox{Hz}^{-1}$. At lower luminosities star-forming galaxies
(SFGs) take over. Luminosities of SFGs reach only $P_{1.4\rm GHz}\simeq
10^{24}\,\hbox{W}\,\hbox{Hz}^{-1}$.

The radio emission of SFGs has two components, both related to star formation:
synchrotron radiation from relativistic electrons and free-free emission from
H\,II regions. The synchrotron emission results from relativistic plasma,
thought to be accelerated primarily in supernova remnants (SNRs) associated
with massive (mass larger than about $8\,M_\odot$) stars that end as Type II
and Type Ib supernovae. Such massive stars live $\le 3\times 10^7\,$yr, and the
relativistic electrons probably have lifetimes $\le 10^8\,$yr
\citep{Condon1992}. Thus the relativistic electrons emitting the  synchrotron
radiation have propagated significant distances ($\ge 1\,$kpc) from their
short-lived ($\ge 10^5\,$yr) and now defunct parent SNRs. Consequently the
original sources of the relativistic electrons have disappeared, and their
detailed spatial distribution has been smoothed beyond recognition
\citep{Condon1992}. Moreover, the steps between star formation and synchrotron
emission (supernova explosion, acceleration of relativistic electrons in the
SNR, propagation of cosmic rays throughout the galaxy, energy loss, and escape)
are poorly understood, impeding a quantitative interpretation of the observed
synchrotron spectra and brightness distributions.

However there is a remarkably tight and ubiquitous correlation between the
global FIR and the (predominantly synchrotron) radio luminosities of star
forming galaxies \citep[e.g.,][]{Yun2001}. Except for galaxies with very low
SFRs, the FIR luminosity appears to be a good measure of the bolometric
luminosity produced by  massive ($M \ge 5\,M_\odot$ ) young stars which are the
most efficient dust heaters. Hence, also the synchrotron luminosity can be
safely exploited as a measure of the SFR. Excluding objects with radio emission
of nuclear origin, star forming galaxies lie very close to a linear relation
between radio and FIR flux densities, with an rms scatter of less than 0.26 dex
\citep{Yun2001}. The radio-FIR correlation was established for local galaxies,
using IRAS data. However, more recent data, primarily from the
\textit{Herschel} observatory, have shown that it persists up to high redshifts
\citep{Ivison2010, Magnelli2015, Delhaize2017, Barger2012, Barger2017}.

The massive stars ionize the H\,II regions as well, powering the free-free
emission whose intensity is therefore proportional to the production rate of
Lyman continuum photons. Thus, the free-free emission:
\begin{itemize}
\item allows the mapping of star-formation regions even in deeply
    dust-enshrouded regions;

\item is a measure of the instantaneous star formation rate, while the
    information from the synchrotron is delayed by $\ge 10^7\,$yr.
\end{itemize}
Radio maps can be made with sub-arcsec position accuracy and resolution,
unambiguously identifying the most luminous star-forming regions within
galaxies and resolving even the most compact ones. At radio  wavelengths also
the densest star-forming regions, generally associated to the most    intense
starbursts, are transparent, so that observed flux densities are accurately
proportional to intrinsic luminosities. Radio stars are rare; therefore their
contamination of emission from star-forming regions is  generally negligible.

Synchrotron, with a spectral index $\alpha_{\rm sync} \sim -0.75$ ($S\propto
\nu^\alpha$), generally dominates at rest-frame frequencies $\simlt
22$--30\,GHz, and above $\sim 100$--150\,GHz the radio emission is overwhelmed
by re-radiation of thermal dust, heated by young stars (cf.
Fig.~\ref{fig:SED_SFG}). Hence, the free-free emission, which has an effective
spectral index at radio frequencies $\alpha_{\rm ff} \sim -0.1$, emerges only
over a narrow frequency range: isolating it and measuring its flux density is
observationally difficult. Note however the timescales. Most of the synchrotron
radiation in a typical galaxy arises from fairly old ($\simgt 10^7\,$yr)
relativistic electrons. Hence, the radio emission in the youngest galaxies
should be free-free. This is expected to be particularly relevant at the
highest redshifts, when cosmic timescales are short.

A calibration of the \textit{SFR-synchrotron emission} relation was obtained by
\citet{Murphy2011} using Starburst99 \citep{Leitherer1999} for a specific
choice of the stellar Initial Mass Function, of the metallicity and of the star
formation history. \citet{Mancuso2015} have slightly modified the relation
including a steepening of the synchrotron spectrum by $\Delta \alpha=0.5$ above
a break frequency of 20 GHz, to take into account  electron ageing effects
\citep{BandayWolfendale1991}. The SFR-synchrotron luminosity relation then
writes:
\begin{equation}\label{eq:murphy}
\bar{L}_{\rm sync}\simeq 1.9\times 10^{21} \left(\frac{\hbox{SFR}}{\hbox{M}_{\odot}\hbox{yr}^{-1}}\right)
\left(\frac{\nu}{\hbox{GHz}}\right)^{-0.85}\left[1+\left({\nu\over 20\rm GHz}\right)^{0.5}\right]^{-1}\!\!\!\! \hbox{W}\,\hbox{Hz}^{-1}.
\end{equation}
This calibration implies a linear relation between $L_{\rm sync}$ and SFR at
all luminosities. However, it has long been suggested that the synchrotron
radiation from low SFR galaxies is somewhat suppressed \citep{Klein1984,
ChiWolfendale1990, PriceDuric1992}, although this view was controversial
\citep{Condon1992}.

A convincing argument in this direction was made by \citet{Bell2003}. He
pointed out that the FIR emission traces most of the SFR in luminous
star-forming galaxies but only a minor fraction of it in low luminosity ones,
as demonstrated by the fact that they are mostly blue, implying that only a
minor fraction of the light from young stars is absorbed by dust. Nevertheless
the FIR to radio luminosity ratio is similar for the two galaxy groups,
implying that the radio emission from low-luminosity galaxies is substantially
suppressed, compared to brighter galaxies. A plausible physical reason for the
suppression of synchrotron emission of low luminosity galaxies is that they are
unable to confine relativistic electrons, which can then escape into the
intergalactic medium before releasing their energy via synchrotron emission.

\begin{figure}
\vskip-0.85cm
\begin{center}
\includegraphics[width=0.7\textwidth]{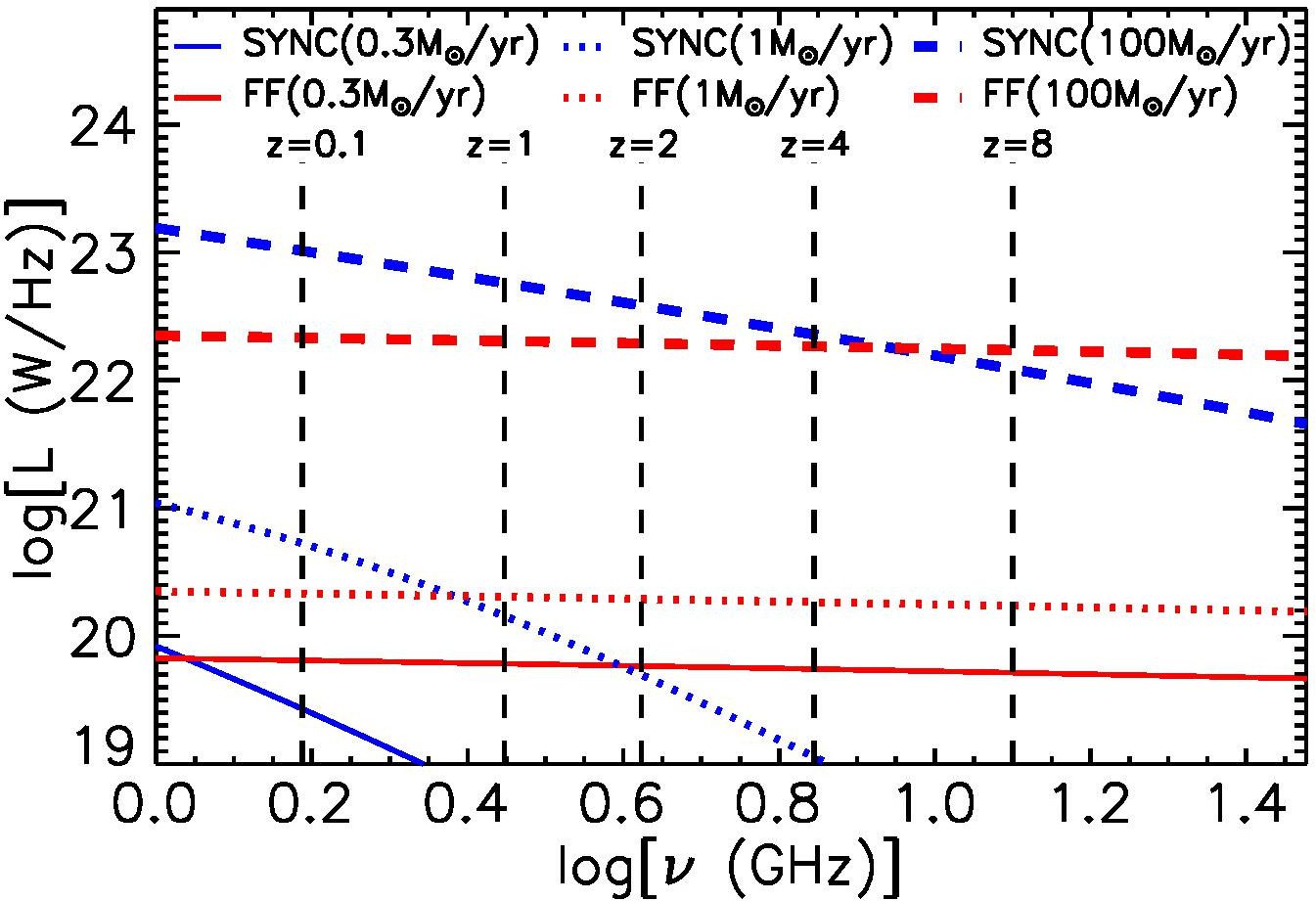}
\end{center}
\vskip-0.5cm
\caption{Synchrotron and free-free luminosity of star-forming galaxies as a function of frequency (in the source frame)
for 3 values of the SFR (0.3, 1 and $100\,M_\odot\,\hbox{yr}^{-1}$). The vertical dashed lines
show  the source-frame frequency corresponding to 1.4 GHz in the observer’s frame.
Objects with low SFRs are expected to be detectable via their free-free emission, particularly at high $z$.
}
\label{fig:sync_ff}
\end{figure}

A direct test of the $L_{\rm sync}$ vs SFR relation can be made comparing the
observationally determined local SFR function \citep{Mancuso2015, Aversa2015}
with the local radio luminosity function \citep{MauchSadler2007}. To reconcile
the SFR function with the radio LF we need to assume that the $L_{\rm
sync}$/SFR ratio declines at low SFRs. Following \citet{Massardi2010},
\citet{Mancuso2015} adopted a relation of the form:
\begin{equation}\label{eq:synch}
L_{\rm sync}(\nu)=\frac{L_{\star,\rm sync}(\nu)}{\left({L_{\star,\rm sync}}/{\bar{L}_{\rm sync}}\right)^{\beta}+
\left({L_{\star,\rm sync}}/{\bar{L}_{\rm sync}}\right)},
\end{equation}
where $\bar{L}_{\rm sync}$ is given by eq.~(\ref{eq:murphy}) and $\beta = 3$.
This equation converges to the \citet{Murphy2011} relation for $\bar{L}_{\rm
sync}\gg L_{\star,\rm sync} = 0.886 \bar{L}_{\rm sync}({\rm
SFR}=1\,M_\odot\,\hbox{yr}^{-1})$. At 1.4\,GHz, $L_{\star,\rm sync}\simeq
10^{28}\, \hbox{erg}\,\hbox{s}^{-1}\,\hbox{Hz}^{-1}$. The agreement at high
luminosities is recovered taking into account a dispersion of $\sigma_{\rm
radio}=0.4\,$dex around the mean  $L_{\rm radio}$--SFR relation.

The suppression of the synchrotron luminosity of low-SFR galaxies make their
free-free emission detectable over a broader frequency range, particularly at
high $z$, as illustrated by Fig.~\ref{fig:sync_ff}. The relation between SFR
and \textit{free-free emission} derived by \citet{Murphy2012}, again using the
Starburst99 stellar population models, writes:
\begin{equation}\label{eq:ff}
 L_{\rm ff}=2.18\times 10^{20} \left(\frac{\hbox{SFR}}{M_\odot/\hbox{yr}}\right) \,
 \left(\frac{T}{10^4\,\hbox{K}}\right)^{0.45}\, \nu_{\rm GHz}^{-0.1}\,\exp{\left(-{h\nu\over k\hbox{T}}\right)}\,\hbox{W}\,\hbox{Hz}^{-1},
\end{equation}
where $T$ is the temperature of the emitting plasma, $h$ is the Planck constant
and $k$ is the Boltzmann constant.
The free-free luminosity at given SFR and for $\hbox{T}=10^4\,$K has thus an
amplitude at 1\,GHz about a factor of 7 lower than the synchrotron luminosity.
However its flatter slope, extending up to $\nu \sim kT/h\simeq 2.1\times
10^{14}\,$Hz, i.e. up to the optical band, implies  $L_{\rm ff}> L_{\rm sync}$
for $\nu \gsim 10\,$GHz, if indeed the synchrotron spectrum steepens above
$\gsim 20\,$GHz.

Since the most extreme starbursts at high redshifts reach SFRs of thousands
$M_\odot/\hbox{yr}$, the highest luminosities at 1.4\,GHz due to star formation
are of several times $10^{24}\,\hbox{W}\,\hbox{Hz}^{-1}$ consistent with the
local luminosity function of \citet{MauchSadler2007}.

The tight relationship between radio and FIR luminosity is exploited as one of
the main indicators to separate radio AGNs from SFGs in deep radio surveys:
excess radio emission with respect to that expected given the SFR is attributed
to nuclear radio activity, leading to a classification as radio AGN. The
correlation is usually parameterized through the so-called $q$ parameter
\citep{Helou1985}:
\begin{equation}
q_{\rm FIR}=\log\left({L_{\rm FIR}[\hbox{W}]\over 3.75\times 10^{12}}\right)-
\log\left(L_{1.4\,\rm GHz} [\hbox{W}\,\hbox{Hz}^{-1}]\right),
\end{equation}
where $L_{\rm FIR}$ is the FIR luminosity integrated from rest-frame 42 to
$122\,\mu$m and $L_{1.4\,\rm GHz}$ is the rest-frame 1.4\,GHz luminosity.

A better choice would be $q_{\rm IR}$, where the FIR luminosity is replaced by
the total (8--$1000\,\mu$m) IR luminosity. However, to compute $L_{\rm IR}$ we
need \textit{Herschel} data, which are available for a small fraction of faint
radio sources. Frequently we need to make do with \textit{Spitzer} $24\,\mu$m
data, a very uncertain proxy of the FIR/IR luminosity. Also, the dividing line
between SFGs and radio-excess sources (RL AGNs) is somewhat arbitrary.

Several models predict that the synchrotron emission is suppressed at high $z$
because of the energy losses of relativistic electrons due to inverse Compton
scattering off the Cosmic Microwave Background photons, whose energy density
increases as $(1+z)^{4}$ \citep[e.g.,][]{Norris2013}. On the contrary,
\citet{Magnelli2015}, based on FIR and radio observations of the most
extensively studied extragalactic fields (GOODS-N, GOODS-S, ECDFS, and COSMOS),
reported evidence of a weak redshift evolution of the parameter $q_{\rm FIR}$,
in the sense of an increasing radio excess with increasing redshift. They
found:
\begin{equation}
q_{\rm FIR}=(2.35 \pm 0.08)\times(1+z)^{\alpha_M},
\end{equation}
with $\alpha_M = -0.12 \pm 0.04$. The weak but statistically significant trend
of $q_{\rm FIR}$ with redshift has been recently confirmed by
\citet{CalistroRivera2017} up to $z\simeq 2.5$ and by  \citet{Delhaize2017} up
to $z \simeq 5$. \citet{Molnar2018} repeated the analysis on the
\citet{Delhaize2017} sample for disc- and spheroid-dominated galaxies
separately. They measured very little change in $q_{\rm  FIR}$ up to $z\simeq
1$--1.5 for disc-dominated galaxies and an increased radio excess with
increasing $z$ at $z\simgt 0.8$ for spheroid-dominated galaxies. They suggested
that this may hint at some residual AGN activity in the latter objects.

The evolution of $q_{\rm FIR}$ reported by \citet{Magnelli2015} implies an
increase with redshift of the $L_{\rm synch}/\hbox{SFR}$ ratio:
\begin{equation}
\log[L_{\rm synch,1.4GHz}(z)]=\log[L_{\rm synch,1.4GHz}(0)]+2.35[1-(1+z)^{-0.12}].
\end{equation}
This is in contradiction with expectations from inverse Compton cooling of
relativistic electrons at high redshifts. On the other hand, a decrease in
$q_{\rm FIR}$ with redshift was expected by some of the most up-to-date
theoretical models \citep[e.g.,][]{Lacki2010, LackiThompson2010,
SchleicherBeck2013} because of the increase with redshift of the SFR surface
density and of the gas density that may better confine relativistic electrons;
the magnetic field may also be amplified.

However, the precise mechanism accounting for the radio--FIR correlation,
including the evolution of $q_{\rm FIR}$ is not well understood yet. A direct
test of the $L_{\rm sync}$--SFR relation, made comparing the observational
determinations of the SFR function and of the radio luminosity function, now
available up to high redshifts, has been carried out by \citet{Bonato2017}.

Determinations of the SFR function need to take into account both the
unobscured and the dust-obscured star formation. At high SFRs, the star
formation is almost completely dust-obscured and can be measured via FIR/sub-mm
observations. At low SFR, most star-formation is unobscured and can be measured
via optical/UV observations both in the continuum or in emission lines
\citep[primarily Ly$\alpha$ and H$\alpha$;][]{Cai2014, Aversa2015}.

The radio luminosity functions at 1.4\,GHz derived from the SFR functions via
eqs.~(\ref{eq:synch})  and (\ref{eq:ff}) were found to match quite well the
observational determinations available at several redshifts, up to $z\simeq 5$.
The fits were similarly good both with and without the weak evolution of
$q_{\rm FIR}$ found by \citet{Magnelli2015}. The evolution is preferred by data
at the highest redshifts, which however have larger uncertainties.

Stronger support to the case for an increase with $z$ of the $L_{\rm
synch}/\hbox{SFR}$ ratio is provided by the source counts: without evolution
the predicted counts lie below the observational determinations while the
evolution by \citet{Magnelli2015} leads to good agreement. The counts turn out
to be quite sensitive to the evolution of the $L_{\rm synch}/\hbox{SFR}$ ratio:
a slightly stronger evolution, with the coefficients of the Magnelli relation
at their $1\,\sigma$ limits, yields counts at hundreds of $\mu$Jy flux
densities at the upper limits of observational determinations.

\section{Identification of the faint (sub-mJy) population}\label{sect:identifications}

Determining the nature of the sub-mJy radio sources requires optical/IR
spectroscopy to measure the redshift, hence the luminosity, and to classify
them as SFGs or AGNs. These objects however are very faint in the optical. For
example, the median $R$ magnitude for the Extended Chandra Deep Field-South
(E-CDFS) VLA sample, which reaches $S_{\rm 1.4\,GHz}\simeq 32.5\,\mu$Jy, is
$\sim 23$; and this refers only to sources detected in the $R$ band, while
$\sim 20\%$ of the objects have only an IR counterpart \citep{Bonzini2012}.

Getting spectra for such faint sources is very time consuming (prohibitively so
for the very faint tail) but can in principle be done. Even if we had optical
spectra for all the E-CDFS sources, for faint counterparts one can only see a
couple of lines. This is enough to get a redshift but not to properly classify
the objects \citep{Padovani2016}. The classification  thus turns out to be
quite complex and has to resort to indicators of varying effectiveness and
reliability.

The X-ray emission is a powerful indicator of nuclear activity, but the mere
presence of an active nucleus does not necessarily imply that it is
contributing to the radio emission. If nuclear activity is detected but the
radio luminosity is consistent with the expectation from the SFR of the host
galaxy, the source is classified as a radio quiet (RQ) AGN. This classification
criterion has at least a couple of limitations. First, the minimum luminosity
detectable by X-ray surveys increases with increasing redshift and even the
deepest currently available surveys miss a large fraction of high-$z$ AGNs
\citep[cf. Fig.~1 of][]{Bonzini2013}. Second, X-ray luminosities of up to $\sim
10^{42}\,\hbox{erg}\,\hbox{s}^{-1}$ can be accounted for by processes related
to star formation and this prevents a firm identification of weak nuclear
activity.

Whenever X-ray data are missing, the identification of RQ AGNs can be made
using deep observations with the Spitzer Infrared Array Camera (IRAC) at 3.6,
4.5, 5.8 and $8\,\mu$m. Different extragalactic sources occupy somewhat
different regions in the diagram of the $S_{8.0}/S_{4.5}$ versus
$S_{5.8}/S_{3.6}$ ratios \citep[][and references therein]{Donley2012}. This is
because the AGN emission can heat up the surrounding dust that re-emits this
energy in the mid-IR. If the AGN is sufficiently luminous compared to its host
galaxy, the emission from the dust heated by it (which is warmer than that
heated by stars) can produce a power-law thermal continuum across the four IRAC
bands. Sources with this spectral shape occupy a specific region in the IRAC
colour–colour diagram, the so-called Lacy wedge \citep{Lacy2004}. The
completeness of this selection method is therefore high ($\sim 75\%$) at $L_x
\ge 10^{44}\,\hbox{erg}\,\hbox{s}^{-1}$,  but relatively low ($\simlt 20\%$)
for $L_x \le 10^{43}\,\hbox{erg}\,\hbox{s}^{-1}$. The IRAC selection appears
also to be incomplete to radio galaxies \citep{Donley2012} and also easily
misses Type 2 Seyfert galaxies. Nevertheless, the mid-IR (MIR) selection is
very important since it identifies also heavily obscured AGNs (provided that
they are not hosted by a particularly bright galaxy), many of which are missed
even by deep X-ray surveys \citep{Donley2012}.

In short, to classify faint radio sources:
\begin{itemize}
\item one first selects radio AGNs using a variant of the IR--radio
    correlation;

\item then radio AGNs are separated from SFGs using X-ray observations;

\item next, the IRAC colour-colour diagram or SED (spectral energy
    distribution) fitting decomposition \citep{Delvecchio2017} are used to
    recover (RQ) AGNs missed by the X-ray criterion;

\item finally, other indicators (radio spectra, radio morphology, optical
    lines, optical photometry, …), if available, are applied to catch
    possible outliers.

\end{itemize}
The classification work has demonstrated that early attempts overestimated the
fraction of SFGs among sub-mJy sources. This was caused by a selection effect:
SFGs have higher optical to radio luminosity ratios than RL AGNs and therefore
are over-represented in samples identified through magnitude limited  optical
imaging. Only $\simeq 44\%$ of the complete sample of sub-mm radio sources
studied by \citet{Windhorst1985} were optically identified, and $\simeq 70\%$
of these were found to be galaxies, mostly blue (star forming). But this fact
did not warrant that the majority of sub-mJy sources are star-forming galaxies,
as assumed by several authors.

The most recent, extensive study of the composition of the faint radio
population \citep{Smolcic2017} was carried out on the sample from the Very
Large Array Cosmic Evolution Survey (VLA-COSMOS) 3 GHz Large Project. The
survey covers a $2.6\,\hbox{deg}^2$ area with a mean rms of $\sim
2.3\,\mu$Jy/beam. The radio data were combined with optical, near-infrared
(UltraVISTA), and mid-infrared (Spitzer/IRAC) data, as well as X-ray data
(Chandra). This yielded counterparts for $\simeq 93\%$ of the total radio
sample of 10,830 sources, reaching out to $z \simlt 6$.

\citet{Smolcic2017} first selected {moderate-to-high radiative luminosity
 AGN (HLAGN)} by using a combination of X-ray  ($L_\mathrm{X}>10^{42}$ erg/s), IRAC colour-colour  and
 SED-fitting  criteria. About 30\% of the HLAGN show a
 $>3\sigma$ radio-excess in $\log {( L_\mathrm{1.4GHz} / \mathrm{SFR} )}$ diagram,
 while  the remaining $\sim$70\% have a radio luminosity consistent (within
 $\pm3\sigma$) with the star-formation rates in their host galaxies.

SFGs were selected based on evidence of star formation (either blue and green
near-UV (NUV)-optical colours, or red colours but with a detection by
\textit{Herschel}) and excluding objects with $>3\sigma$ radio-excess. Finally,
low-to-moderate radiative luminosity AGN (MLAGN) were drawn from the sample
remaining after exclusion of the HLAGN, selecting sources with $>3\sigma$
radio-excess (radio-excess-MLAGN) and sources with red NUV-optical colours and
no \textit{Herschel} detection (quiescent-MLAGN).

The contribution of MLAGN to the 3\,GHz source counts decreases from about 75\%
at $S_\mathrm{3GHz}\sim400-800~\mu$Jy to about 50\% at
$S_\mathrm{3GHz}\sim100-400\,\mu$Jy, and  to $\simeq 20\%$  at
$S_\mathrm{3GHz}\sim50\,\mu$Jy. In the same flux density interval, the fraction
of SFGs increases from $\simeq 10\%$ ($S_\mathrm{3GHz}\sim400-800~\mu$Jy) to
$\simeq 60\%$ ($S_\mathrm{3GHz}\sim50~\mu$Jy). The fraction of HLAGN remains
approximately constant, at 20-30\%. Hence the combined AGN sample (MLAGN and
HLAGN) dominates at $S_\mathrm{3GHz}\simgt100~\mu$Jy while SFGs take over at
lower flux densities.

The above definitions of MLAGNs and HLAGN imply that a fraction of them have
radio emission consistent with that expected from star formation. Separating
the sources into two groups, with and without $>3\sigma$ radio excess,
\citet{Smolcic2017} found that the fraction of sources whose radio luminosity
can be accounted for by star formation increases from about 10\% at
$S_\mathrm{1.4GHz}\sim700-1000~\mu$Jy to  $\sim 50\%$ at
$S_\mathrm{1.4GHz}\sim200ù,\mu$Jy, to  $\sim85\%$ at
$S_\mathrm{1.4GHz}\sim50~\mu$Jy.

Similar results, once we take into account the differences in the selection
criteria, were reported by \citet{Bonzini2013} who classified the sources
detected in the 1.4\,GHz VLA survey  of the Extended Chandra Deep Field-South
(E-CDFS). The VLA/E-CDFS survey covered an area of $0.3\,\hbox{deg}^2$ reaching
a flux density limit of $\simeq 32\,\mu$Jy.

\section{Radio quiet (RQ) AGNs: a new radio source population?}\label{sect:RQ}

Although the discovery of quasi stellar objects (QSOs)\footnote{Throughout this
paper we will use preferentially the term AGN to refer to sources powered by
nuclear activity. However, AGNs will also be referred to as ``QSOs'' or
``quasars'' when these terms are used in the cited papers.} sprang from the
optical identification and redshift measurement of a powerful radio source
\citep[3C273;][]{Schmidt1963} and the first QSOs were radio-selected, it became
clear relatively soon \citep{Sandage1965} that the majority of QSOs are not
radio loud (RL);  they are usually referred to as radio quiet (RQ). Several
follow-up surveys at radio frequencies of optically selected QSOs, reviewed by
\citet{Stern2000a}, typically detected between 10\% and 40\% of them.

The fraction  of radio-detected QSOs increased with the depth of radio surveys,
suggesting that RQ AGNs are perhaps only “radio-faint”. As we have seen,
identifications of sub-mJy radio sources have revealed the presence of a
substantial fraction RQ  AGNs.  Evidence of nuclear activity in these sources
comes from other bands (e.g. optical, mid-infrared, X-ray) so that the link
between the nuclear activity and the radio emission is unclear and is still
being hotly debated.

Many papers have addressed the question: is there a continuity between RL and
RQ AGNs, so that the latter simply populate the low radio luminosity tail of
the AGN distribution or RQ and RL AGNs are two totally distinct populations, at
least as far as radio properties are concerned?

\citet{Kellermann1989} reported a bimodal distribution of the radio flux
density of QSOs from the optically selected Palomar Bright Quasar survey
observed with the Very Large Array (VLA) at 5 GHz. The bimodality suggests that
two distinct physical processes are present, with one process being
significantly more powerful than the other for different QSO classes. Deep
Expanded VLA (EVLA) observations at 6 GHz with nearly complete (97\%) radio
detections in a volume-limited colour-selected sample of 179 QSOs
\citep{Kimball2011} have shown that the radio luminosity function (LF) can be
explained as a superposition of two populations: QSO radio sources with
$\log(L_{6\,\rm GHz}/\hbox{erg}\, \hbox{s}^{-1}\, \hbox{Hz}^{-1})> 30$ are
powered primarily by nuclear activity, while those with lower 6\,GHz luminosity
are dominated by star formation in the QSO host galaxies. Support to this
conclusion was provided by \citet{Condon2013} who used the 1.4 GHz NRAO VLA Sky
Survey (NVSS) to study radio sources in two colour-selected QSO samples.

A careful analysis  of the dichotomy in the radio loudness distribution of QSOs
was carried out by \citet{Balokovic2012} by modeling their radio emission and
various selection effects using a Monte Carlo approach. The resulting simulated
samples were compared to a fiducial sample of 8300 QSOs drawn from the seventh
data release of the Sloan Digital Sky Survey (SDSS DR7) Quasar Catalog and
combined with radio observations from the Faint Images of the Radio Sky at
Twenty cm (FIRST) survey. Their results indicated that the SDSS--FIRST sample
is best described by a radio loudness distribution consisting of two
components, with ($12\pm 1)$\% of sources in the radio-loud component. The
evidence for bimodality was however found to be not strong.

Based on their study of the  Extended Chandra Deep Field-South (E-CDFS) VLA
sample \citet{Padovani2015} concluded that RQ and RL QSOs are two totally
distinct populations,  characterized by very different evolutions, luminosity
functions, and Eddington ratios. The radio power of RQ AGNs evolves similarly
to star-forming galaxies, consistent with their radio emission being powered by
star formation. This conclusion was confirmed by the study of
\citet{Bonzini2015} who used deep \textit{Herschel} photometry to determine the
FIR luminosity, hence the SFR, of E-CDFS galaxies.

On the other hand,  \citet{Lacy2001} and \citet{Cirasuolo2003} found a
continuous distribution of radio-to-optical flux ratios, without a minimum
between those of RL and RQ AGNs \citep[but see ][for a different
interpretation]{Ivezic2002, Ivezic2004}. This view was corroborated by
\citet{Barvainis2005} who,  through a study of radio variability across a broad
range of radio power, from quiet to loud, came to the conclusion that ``... the
radio emission from radio-quiet quasars originates in a compact structure
intimately associated with the active nucleus. The alternative hypothesis, that
the emission from radio-weak quasars is from a starburst, is ruled out.''

A similar conclusion was reached by \cite{Bonchi2013} studying the dependence
of the nuclear radio (1.4\,GHz) luminosity on both the 2--10 keV X-ray and the
host-galaxy K-band luminosities for a complete sample of 1268 X-ray-selected
AGNs. No evidence of bimodality in the radio luminosity distribution was found.
This continuity is consistent with a nuclear origin of the radio emission for
both RQ and RL AGNs.

\citet{White2015} used a combination of optical  and near-infrared photometry
to select, from the VISTA Deep Extragalactic Observations (VIDEO) survey, a
sample of 74 objects that can be confidently classified as QSOs (`Gold
candidate quasar sample')  over $1\,\hbox{deg}^2$. By comparing the radio
flux-density distribution for these QSOs  with that for a control sample of
galaxies matched in redshift and stellar mass, and by estimating the star
formation rate, they found that the QSOs have excess radio flux when assuming
the most reasonable values of black hole masses. This result was interpreted as
indicating ``that accretion is the primary origin of the quasars’ total radio
emission.''

Direct evidences of a nuclear origin of the radio  emission from RQ AGNs were
provided via high angular resolution observations by
\citet{GirolettiPanessa2009},  \citet{Guidetti2013}, \citet{Jackson2015} and
\citet{HerreraRuiz2016}. \citet{Zakamska2016} assembled two samples of
optically obscured or unobscured AGNs for which the host star formation rates
(SFRs) could be estimated or constrained with infrared spectroscopy and
photometry from \textit{Spitzer} and \textit{Herschel}. The SFRs were found to
be insufficient, by an order of magnitude, to explain the observed radio
emission derived by cross--matching the samples with the FIRST survey catalog.
They concluded that, although RQ AGNs in their samples lie close to the
infrared/radio correlation characteristic of the star-forming galaxies, both
their infrared emission and their radio emission are dominated by the nuclear
activity, not by the host galaxy.

It is clear from the above review that the radio properties of RQ AGNs are not
well understood yet. Observational indications both of a nuclear and of a star
formation origin have been reported. The two mechanisms frequently co-exist and
it is unclear which is the dominant one, in a statistical sense. The galaxy-AGN
co-evolution model by \citet{Cai2013}, extended by \citet{Cai2014},
\citet{Bonato2014} and \citet{Mancuso2015}, allows us to investigate whether or
not the amount of star formation of AGN hosts is sufficient to explain the
counts and the radio luminosity functions of RQ AGNs associated to sub-mJy
radio sources.

\begin{figure*}
\includegraphics[width=\textwidth]{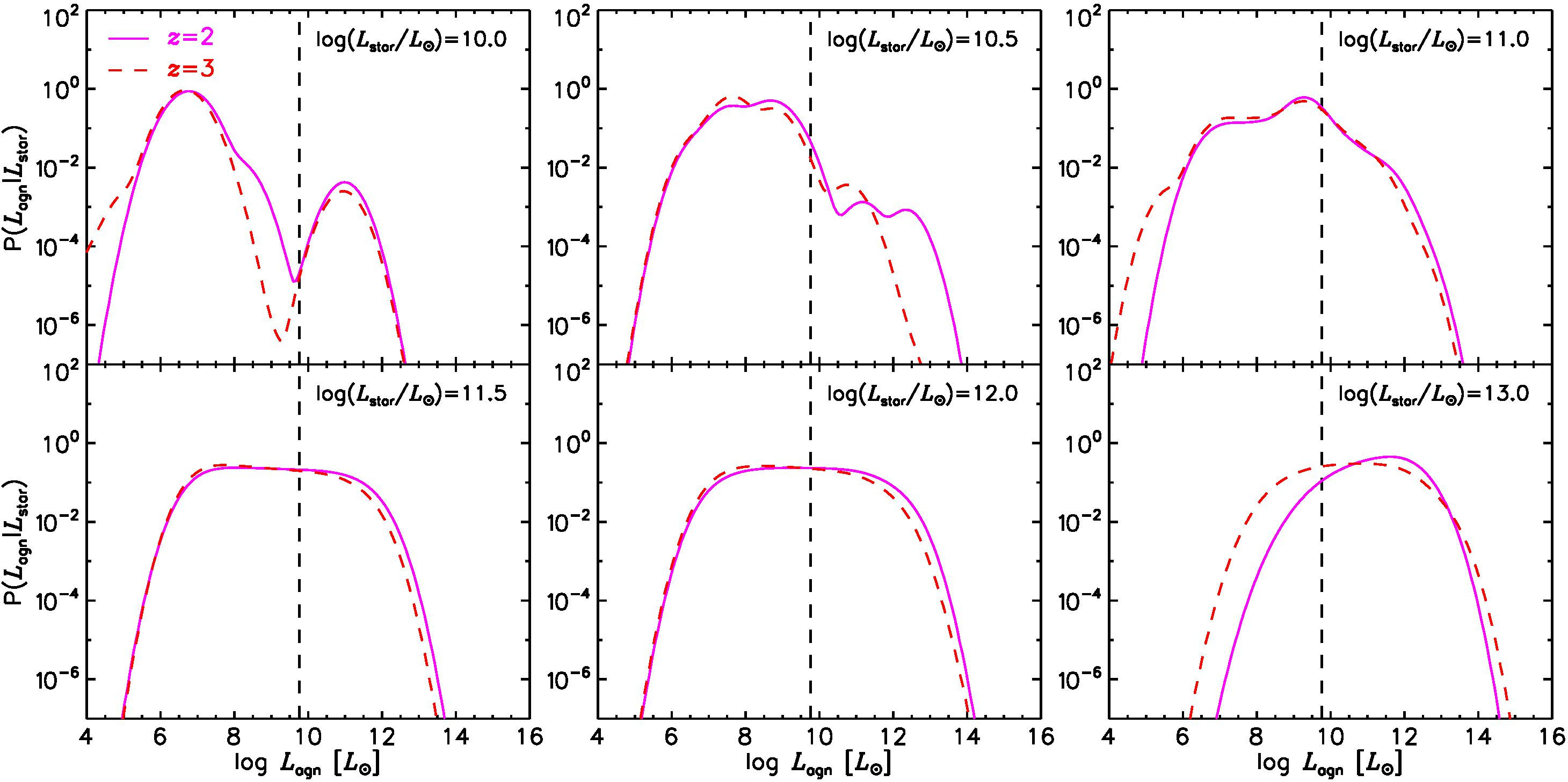}
\caption{Probability that a star-forming proto-spheroidal galaxy with luminosity
$L_{\rm star}$ at redshift $z=2$ or $z=3$ (solid and dashed lines, respectively)
hosts an AGN with luminosity $L_{\rm agn}$ within $d\log L_{\rm agn}$
[eq.~(\ref{eq:mayberight})]. $L_{\rm star}$ is a measure of the SFR [eq.~(\ref{eq:calib_spher})].
The vertical dashed lines correspond to
$L_{\rm agn}=22.4\times 10^{42}\, \hbox{erg}\, \hbox{s}^{-1}=5.8\times 10^9\,L_\odot$,
where $10^{42}\, \hbox{erg}\, \hbox{s}^{-1}$ is the minimum X-ray luminosity of
E-CDFS VLA sources classified as RQ AGNs by \citet{Bonzini2013} and 22.4 is the
bolometric correction adopted by \citet{Chen2013}.}
 \label{fig:prob2}
\end{figure*}

The model comprises two populations of star-forming galaxies, late-types that
are dominating the cosmic star formation rates at $z\simlt 1.5$, and
proto-spheroids, that formed the bulk of their stars at higher redshifts in a
relatively short time, $\simlt 1\,$Gyr, at $z> 1$--1.5.

The formation  and evolution of proto-spheroidal galaxies is linked to the
formation of massive halos. The baryonic gas falling into the halo potential
wells is shock-heated to the virial temperature and then cools down giving rise
to star formation. At the same time it loses angular momentum, e.g. by effect
of the radiation drag, and flows towards the central regions at a rate
proportional to the SFR. It is temporarily stored in a reservoir/proto-torus
from which it flows into the super-massive black hole at the rate allowed by
the viscous dissipation of the angular momentum.  A small fraction of the
emitted power comes out in mechanical form. In massive objects, this AGN
feedback exceeds that from supernovae (also taken into account by the model)
and the gravitational binding energy of the gas. The gas is thus eventually
swept out of the potential well. Consequently both the star formation and the
gas supply to the reservoir are halted. The nuclear activity however continues
until the reservoir is depleted.

These processes are  governed by a set of equations, laid down in the Appendix
of \citet{Cai2013}. The formalism has allowed the calculation of the evolution
with galactic age of the SFR and of the AGN bolometric luminosity for any
galaxy halo mass and redshift. We can then compute the probability that a
galaxy of given redshift and given SFR hosts an AGN with X-ray luminosity
$L_\mathrm{X}>10^{42}\,\hbox{erg}\,\hbox{s}^{-1}$ and is therefore classified
as a RQ AGN.

The probability that a galaxy with total luminosity due to star formation
$L_{\rm star}$ at redshift $z$ hosts an AGN with bolometric luminosity $L_{\rm
agn}$ within $d\log L_{\rm agn}$, hence with total luminosity $L_{\rm
tot}=L_{\rm star}+L_{\rm agn}$, is:
\begin{equation}\label{eq:mayberight}
	P_{\rm protosph}(L_{\rm agn}|L_{\rm star}, z) \equiv {\Phi_{\rm agn}[L_{\rm agn}|L_{\rm tot}, z]
\over {\mathlarger\int}^{\log L_{{\rm agn},\rm max}}_{\log L_{{\rm agn},\rm min}}
\Phi_{\rm agn}[L_{\rm agn}|L_{\rm tot}, z] d\log L_{\rm agn}},
\end{equation}
where $\Phi_{\rm agn}(L_{\rm agn}|L_{\rm tot}, z)$ is  the conditional AGN
luminosity function per unit $d\log L_{\rm agn}$, derived by
\citet{Bonato2014}:
\begin{eqnarray}\label{eq:prob6}
&& \Phi_{{\rm agn}}(L_{{\rm agn}}\vert L_{\rm tot},z) = \int_{M_{\rm min}}^{M_{\rm max}} dM_{\rm vir} \nonumber \\
&& \times \int_{z}^{z_{\rm max}} dz_{\rm vir} \bigg\vert \frac{dt_{\rm vir}}{dz_{\rm vir}} \bigg\vert \frac{dN_{\rm ST}}{dt_{\rm vir}} (M_{\rm vir},z_{\rm vir}) \frac{1}{2\pi\sigma_{{\rm star}}\sigma_{{\rm agn}}} \nonumber \\
&& \times \frac{L_{\rm tot}}{L_{\rm tot}-L_{{\rm agn}}}\exp\{-[\log (L_{\rm tot}-L_{{\rm agn}})-\log \bar{L}_{\rm star}]^{2}/2\sigma_{{\rm star}}^2\}\nonumber \\
&& \times \exp[-(\log L_{{\rm agn}}-\log \bar{L}_{{\rm agn}})^{2}/2\sigma_{{\rm agn}}^2].
\end{eqnarray}
In the above equation $\bar{L}_{\rm star} = \bar{L}_{\rm star}(z\vert M_{\rm
vir},z_{\rm vir})$ and $\bar{L}_{{\rm agn}} = \bar{L}_{{\rm agn}}(z\vert M_{\rm
vir},z_{\rm vir})$ are the redshift-dependent mean {starburst} and AGN
luminosities at given halo mass, $M_{\rm vir}$, and virialization redshift,
$z_{\rm vir}$, respectively, and $\sigma_{{\rm star}}$ and $\sigma_{{\rm agn}}$
are the corresponding dispersions for which we have adopted the values derived
by \citet{Cai2013}: $\sigma_{{\rm star}}=0.1$ and $\sigma_{{\rm agn}}=0.35$.
$dN_{\rm ST}/dt$ is the dark matter halo mass function \citep{ShethTormen1999}.

Examples of the probability distributions given by eq.~(\ref{eq:mayberight})
are shown in Fig.~\ref{fig:prob2} for two values of the redshift ($z=2$ and
$z=3$) and for several values of $L_{\rm star}$, which is a measure of the SFR:
\begin{equation}\label{eq:calib_spher}
\log \Big(\frac{L_{\rm star}}{L_\odot}\Big) =
\log \Big(\frac{{\rm SFR}}{M_\odot\ {\rm yr}^{-1}}\Big) + 9.892.
\end{equation}
As mentioned above, the  local luminosity function by \citet{MauchSadler2007}
can be reproduced adopting a Gaussian distribution of the radio luminosities
with a dispersion $\sigma_{\rm radio}=0.4\,$dex around the mean $L_{\rm
radio}$--SFR relation:
\begin{equation}\label{eq:lum_distr}
\Phi_{\rm radio}(L_\nu|{\rm SFR})= \frac{\phi({\rm SFR})}{\sigma_{\log L}\sqrt{2\pi}}
\exp\left\{{-{1\over 2}\left[{\log[L_\nu/\bar{L}_\nu({\rm SFR})]\over
\sigma_{\log L}}\right]^2}\right\}{d\log({\rm SFR})\over d\log({L}_\nu)},
\end{equation}
where $\bar{L}_\nu({\rm SFR})= L_{\rm sync}(\nu|{\rm SFR})+L_{\rm ff}(\nu|{\rm
SFR})$ and  $\sigma_{\log(L)}=0.4$.

The contribution  of proto-spheroidal galaxies to the radio LF of RQ AGNs is
then obtained multiplying the distribution of radio luminosities of galaxies at
fixed SFR [eq.~(\ref{eq:lum_distr})] by the probability that a galaxy with such
SFR (or with the corresponding total luminosity due to star formation, $L_{\rm
star}$), at redshift $z$ hosts an AGN with bolometric luminosity $L_{\rm agn}>
22.4\times 10^{42}\, \hbox{erg} \, \hbox{s}^{-1}$. Here 22.4 is the bolometric
correction \citep{Chen2013} and $10^{42}\, \hbox{erg} \, \hbox{s}^{-1}$ is the
minimum X-ray luminosity of sources classified as RQ AGNs by
\citet{Bonzini2013}. This probability is computed integrating
eq.~(\ref{eq:mayberight}) over $d\log L_{\rm agn}$.

As for late-type galaxies, \citet{Bonato2014} argued that the probability of
finding a RQ AGN with $L_{\rm X}> 10^{42}\, \hbox{erg} \, \hbox{s}^{-1}$ inside
a star--forming galaxy with a given SFR (or associated luminosity $L_{\rm
star}$) can be be described by a Gaussian distribution  with a dispersion
$\sigma_{\rm lt}=0.69\,\hbox{dex}$ around the mean $\bar{L}_{\rm
X}(\hbox{SFR})$:
\begin{equation}\label{BH mean}
\log \left({\bar{L}_{\rm X}\over \hbox{erg}\,\hbox{s}^{-1}}\right)=
30.17 + 1.05\log\left({L_{\rm IR}\over L_\odot} \right).
\end{equation}
The wanted probability is then:
\begin{equation}
P_{\rm lt}(L_\nu)={1\over \sigma_{\rm lt}\sqrt{2\pi}}
\exp\left\{{-{1\over 2}\left[{\log(L_{\rm X}/\bar{L}_{\rm X}({\rm SFR}))
\over \sigma_{\rm lt}}\right]^2}\right\}{d\log({\rm SFR})\over
d\log(\bar{L}_\nu)}.
\end{equation}
The contribution  of late-type galaxies to the radio LF of RQ AGNs then follows
as in the case of proto-spheroids.

The resulting radio luminosity functions of RQ AGNs turn out to be in good
agreement with observational determinations, available at several redshifts
\citep[][and paper in preparation]{Bonato2017}. This means that the data are
consistent with the hypothesis that the radio emission associated to star
formation can account for the statistical properties (number counts,
redshift-dependent luminosity functions) of faint radio sources not associated
to RL AGNs, with and without nuclear activity detectable in other wavebands.
However, this does not mean that the radio emission of RQ AGNs must be
negligible. The uncertainties are large enough to leave room for significant,
although not dominant, contributions from these objects.

In summary, recent surveys of the faint radio sky are playing an important role
in a variety of astrophysical topics, including the understanding of the cosmic
star formation history, the galaxy evolution, the existence of powerful jets,
and the radio emission in RQ AGNs.

Assessing the nature of sub-mJy radio sources took more than thirty years
because radio surveys were exploring galaxies fainter than those accessible in
other wavebands. Now it is well established that below $S_{1.4\rm GHz} \simeq
0.1\,$mJy the radio sky is dominated by SFGs. As we will see in the next
section, the Square Kilometre Array (SKA) that will reach flux limits orders of
magnitude fainter than is currently possible, and over large areas, will boost
the potential of radio astronomical observations for  extragalactic studies
related to star formation and galaxy and AGN evolution.

Radio observations are unaffected by absorption, which means, for example, that
they can dig out the most extremely dust-enshrouded star-forming regions and
are sensitive to all types of AGN, irrespective of obscuration and orientation
(i.e., Type 1s and Type 2s). But the full exploitation of the SKA potential
requires synergies with other multi-wavelength astronomical such as the
Advanced Telescope for High ENergy Astrophysics \citep[ATHENA;][]{Barcons2015},
the Wide Field Infrared Survey Telescope \citep[WFIRST;][]{Spergel2015}, the
Large Synoptic Survey Telescope \citep[LSST;][]{LSST2017}, the European
Extremely Large Telescope \citep[E-ELT;][]{Liske2012Liske}, the James Webb
Space Telescope \citep[JWST;][]{Gardner2012, Greenhouse2015}, and the Space
Infrared Telescope for Cosmology and Astrophysics
\citep[SPICA;][]{Nakagawa2017}.

\section{The SKA and the study of the cosmic star-formation history}\label{sect:SKA}

The SKA \citep[see, e.g.,][]{Combes2015, Grainge2017}\footnote{See also\\
\url{www.skatelescope.org/wp-content/uploads/2017/02/SKA-Community-Briefing-18Jan2017.pdf}}
is a new technology radio-telescope array, with one square kilometre collecting
surface, orders of magnitude more sensitive and rapid in sky surveys than
present instruments.

The SKA comprises two separate interferometric arrays. In 2012, it was decided
that they will be hosted by two desert sites, chosen to minimize terrestrial
interference, in different continents. The SKA-low telescope, that will operate
between frequencies of 50 and 350\,MHz, will be deployed at the Murchison Radio
Astronomy Observatory, Australia. At the end of Phase 1 it will incorporate a
total of 131,072 antennas, grouped into 512 stations of 256 antennas each. The
majority (296) of these stations will be located in a ``core'' area;  the
remaining 216 will be deployed along three spiral arms to give a maximum
baseline of 65 km.

The SKA-mid telescope will operate between 350\,MHz and 24\,GHz and will be
located in the Karoo desert in South Africa. At the end of Phase\,1 (SKA1) it
will be made up of 133 antennas with 15-m diameter, plus 64 antennas of 13.5-m
diameter being built for the MeerKAT telescope. Again, there will be a central
core of antennas and three spiral arms allowing a maximum baseline of 156 km.
In Phase\,2 (SKA2) the number of SKA-mid dishes will be increased to $\sim
2000$ across 3500\,km of Southern Africa and a major expansion of SKA1-Low
across Western Australia is also foreseen.

A key feature of the SKA is that traditional dish radio telescope designs,
which will be used in the high frequency component, will be complemented  with
aperture arrays. These are made by large numbers of small, fixed antenna
elements coupled to appropriate receiver systems which can be arranged in a
regular or random pattern on the ground. A signal ``beam'' (or field-of-view,
FoV) is formed and steered by combining all the received signals after
appropriate time delays have been introduced to align the phases of the signals
coming from a particular direction. By simultaneously using different sets of
timing delays, this ``beam forming'' can be repeated many times to create
multiple independent beams. This FoV expansion technology will allow
instantaneous imaging of multiple sky regions simultaneously, massively
increasing the telescope survey speed.

The number of useful beams produced, or total FoV, is essentially limited by
signal processing, data communications and computing capacity. With 8
independent beams, the FoV is of $1\,\hbox{deg}^2$ at 1.4\,GHz (21\,cm) and
could, in principle, be extended by up to a factor of 100. For comparison, the
Hubble Space Telescope (HST) FoV is $\sim 1\,\hbox{arcmin}^2$ and the  ALMA FoV
is $\sim 0.25\,\hbox{arcmin}^2$.

The angular resolution also depends on frequency: it is higher/poorer at
higher/lower frequencies. The Phase\,1 goal is to achieve a resolution of $\sim
0.2\,$arcsec at 1.4\,GHz. In Phase\,2 the resolution at this frequency will
improve to 0.01\,arcsec with baselines of $\sim 3000\,$km.

The construction of the SKA will start in 2018. The SKA1 instruments are
expected to be in place on 2024. The construction cost cap for this phase,
inflation adjusted, is of 674.1 million euro. The SKA2 is expected to start in
the mid-2020's. The total estimated SKA construction cost is about 1.5 billion
euros.

The SKA is a world-wide project. Ten countries  (Australia, Canada, China,
India, Italy, Netherlands, New Zealand, South Africa, Sweden and UK) are
currently members of the SKA organization. Several other countries (France,
Germany, Japan, Korea, Malta, Portugal, Spain, Switzerland, USA) have expressed
interest in joining.

The three radio telescopes that have been built on the two selected SKA sites
are called the SKA precursors, to be included in the final instrument. They are
the Murchison Widefield Array (MWA) and ASKAP in Australia, and MeerKAT in
South Africa. The MWA is a low-frequency radio telescope operating between 80
and 300 MHz. It is performing large surveys of the entire Southern sky and
acquiring deep observations on targeted regions.

ASKAP is designed to be a high speed survey instrument with high dynamic range.
It has a total collecting area of approximately $4,000\,\hbox{m}^2$, from 36
antennas each $12\,$m in diameter; covers the frequency range from 700 MHz to
1.8 GHz; operates with 36 independent beams, each of about $1\,\hbox{deg}^2$,
yielding a total FoV of $30\,\hbox{deg}^2$ at 1.4\,GHz.  The ASKAP Early
Science Program started in October 2016 using an array of 12 antennas.

The MeerKAT is currently under construction in South African Karoo region. It
is a precursor to the full SKA system but also operates as an independent
instrument that will be conducting critical science for some years before being
integrated into the SKA1. When completed it will be made of 64 dishes each of
13.5\,m in diameter. The first seven dishes are complete and are known as
KAT-7.

In addition to the SKA precursors, there are the SKA Pathfinders that will not
be part of SKA but contribute to scientific and technical developments  of
direct use to the SKA (e.g., LOFAR).

The SKA science case is extensively discussed in the book ``Advancing
Astrophysics with the Square Kilometre Array'' available at
\url{https://www.skatelescope.org/books/}. The book comprises 135 chapters
written by 1213 contributors. The science spans many diverse areas of
astrophysics, with major advances expected in the following areas:
\begin{itemize}
\item Strong-field tests of gravity with pulsars and black holes.

\item Cosmic dawn and the epoch of reionization.

\item The transient radio sky.

\item Cosmology and dark energy.

\item The origin and evolution of cosmic magnetism.

\item Galaxy evolution probed by neutral hydrogen.

\item The cradle of life and astrobiology.

\item Galaxy and cluster evolution probed in the radio continuum.

\end{itemize}
Studies of the star formation across cosmic time and of the galaxy-AGN
co-evolution are among the top science priorities/drivers for continuum SKA1
surveys. A set of \textit{reference surveys} have been identified
\citep{PrandoniSeymour2015}. Three of them are at $\simeq 1\,$GHz, all with
angular resolution of $0.5\,$arcsec, to avoid reaching the confusion limit:
\begin{itemize}

\item An ultra-deep survey (rms $0.05\,\mu$Jy/beam) over an area of
    $1\,\hbox{deg}^2$.

\item A deep survey (rms $0.2\,\mu$Jy/beam) over an area of
    $10-30\,\hbox{deg}^2$.

\item A wide survey (rms $1\,\mu$Jy/beam) over an area of
    $1000\,\hbox{deg}^2$.

\end{itemize}
According to the detailed predictions by \citet{Mancuso2015}, the ultra-deep
survey will detect, at $5\,\sigma$, high-$z$ galaxies with SFRs two orders of
magnitude lower compared to \textit{Herschel} surveys, that were confusion
limited. This means that SKA1 will reach SFRs of a few
$M_\odot\,\hbox{yr}^{-1}$ at $z\simeq 1$. The minimum detectable SFR increases
with increasing redshift but is still of $\hbox{few}\times
M_\odot\,\hbox{yr}^{-1}$ at $z\simeq 10$, implying that the SKA1 ultra-deep
survey can detect galaxies up to the highest redshifts, $z \simeq 10$.

The maximum redshift of detectable sources decreases to about 8 for the deep
survey and to about 7 for the wide survey. Nevertheless, also these surveys
will dig inside the re-ionization epoch. The highest redshift tails of the
distributions at the detection limits of all these surveys comprise a
substantial fraction of strongly lensed galaxies. \citet{Mancuso2015} predicted
that the ultra-deep survey will detect about 1200 strongly lensed galaxies per
square degree, at redshifts of up to 10. For about 30\% of them the SKA1 will
detect at least 2 images.

Two additional surveys, with a resolution of $0.1\,$arcsec, are foreseen at
$\simeq 10\,$GHz:
\begin{itemize}

\item An ultra-deep survey (rms $0.04\,\mu$Jy/beam) over an area of
    $0.0081\,\hbox{deg}^2$.

\item A deep survey (rms $0.3\,\mu$Jy/beam) over an area of
    $1\,\hbox{deg}^2$.

\end{itemize}
Surveys at this higher frequency have two advantages compared to those at
$\simeq 1\,$GHz:
\begin{itemize}
\item They yield higher angular resolution, allowing us to resolve
    star-forming galaxies at substantial redshifts and to study the detailed
    astrophysics of star formation. They thus provide an extinction-free view
    for the morphologies of dusty star-bursting galaxies. These observations
    enable to distinguish between competing theories advocating either a star
    formation triggered by mergers or regulated by in-situ processes, or
    fuelled by large-scale flows of gas.

\item At this frequency,  emission of star-forming galaxies becomes dominated
    by free-free radiation. As discussed in section~\ref{sect:SFG}, at
    variance with synchrotron emission, the free-free is a measure of the
    instantaneous star formation rate, and allows the mapping of
    star-formation regions even in deeply dust-enshrouded regions.

\end{itemize}
In summary, the planned SKA1 continuum surveys will allow a great progress in
the determination the star formation rate across cosmic times and in the
understanding of the astrophysical processes going on within star forming
galaxies. Also the SKA1 sensitivity and resolution will allow to detect nuclear
activity down to very faint flux density levels, clarifying the radio
properties of RQ AGNs. The large samples down to very faint flux densities
provided by SKA1 surveys will enable a complete census of star-formation and
AGN activity as a function of redshift, of galaxy mass and of galaxy
morphology.

\section{Conclusions}\label{sect:conclusions}

Radio astronomy has been playing a leading role in many fields of astronomical
research. For decades, from the early 1950's to mid 1990's, identification of
radio sources has remained the most effective method for finding high-$z$
galaxies. Radio astronomical observations have led to Nobel prize discoveries
(the CMB and its anisotropies, the pulsars and their timing, allowing tests of
general relativity in a strong gravitational field) as well as to other
discoveries revolutionizing our understanding of the universe: the quasars and
the ensuing understanding of black hole accretion as a powerful energy source,
the cosmological evolution. In the 1950's they pushed the boundary of the
observable universe far beyond the reach of optical telescopes of the time.

Radio source counts down to mJy flux density levels are dominated by radio
AGNs. But a different population, star forming galaxies, is increasingly
important at fainter and fainter flux densities, and becomes dominant below
$\simeq 0.1\,$mJy at 1.4 GHz. Hence, deep radio surveys are providing another
tool to investigate the cosmic star formation history.

So far, radio surveys have played only a marginal role in this respect, but the
SKA and, to some extent, its precursors will change this drastically, pushing
the radio band at the forefront in terms of sensitivity to SFR in comparison to
other bands.

The SKA2 is expected to detect, with a 1,000\,h exposure,  a galaxy with a SFR
of $\sim 10\,M_\odot\,\hbox{yr}^{-1}$ up to $z\sim 8$ \citep{Murphy2015}, thus
providing a more complete view of the cosmic star formation than any other
forthcoming instrument.

With 200 km baselines, observations at $\sim 10\,$GHz will be able to achieve a
maximum angular resolution $\simlt 0.03\,$arcsec, sampling 250\,pc scales
within disk galaxies at $z\sim 1$, providing an extinction-free view for the
morphologies of dusty star-bursting galaxies that dominate the star formation
activity between $1 \simlt z \simlt 3$.

While probing such fine physical scales requires sources that are extremely
bright, the SKA should still have the sensitivity to easily resolve
sub-$L_\star$ galaxies\footnote{$L_\star$ galaxies at $1 \simlt z \simlt 3$
have total (8--$1000\,\mu$m) infrared luminosities in the range $10^{11} \simlt
L_{\rm IR}/L_\odot \simlt 10^{12}$} at high redshifts \citep{Murphy2015}.
SKA1-mid will easily resolve such sources at $z \simlt 2$.

In the same redshift range, galaxies with $L_{\rm IR} \simgt 10^{12}\,L_\odot$
can be \textit{mapped} by the SKA at $\sim 10\,$GHz with a 0.1\,arcsec
resolution. Such resolution matches that of the JWST and enables a view of star
formation activity even in the most dust-enshrouded regions.

These high-sensitivity, high angular resolution observations will shed light on
the debated issue of the radio emission of RQ AGNs, distinguishing between the
contribution from star formation and from AGN activity.

\acknowledgements{GDZ renews his appreciations for the kind invitation Third
Cosmology School and the extraordinarily warm hospitality.  Work supported in
part by ASI/INAF agreement n.~2014-024-R.1 for the {\it Planck} LFI Activity of
Phase E2 and by the ASI/Physics Department of the university of Roma--Tor
Vergata agreement n. 2016-24-H.0 for study activities of the Italian cosmology
community.}

\bibliographystyle{ptapap}
\bibliography{ptapapdoc}

\end{document}